\documentclass[review,12pt,authoryear]{elsarticle}
\usepackage[pagewise]{lineno}
\usepackage{graphicx}
\graphicspath{{C:/Users/Ferran/Documents/UPC/DOCTORAT/Article_Passes2013/Article/Paper_Graphs/} }
\journal{Journal of Atmospheric and Solar-Terrestrial Physics}

\begin{document}
\begin{frontmatter}
\title{Analysis of global Terrestrial Gamma Ray Flashes distribution and special focus on AGILE detections over  South America}

\author[upc]{Ferran Fabr\'o \corref{cor1}}
\ead{ferran.fabro@upc.edu}
\author[upc]{Joan Montany\`a}
\author[inaf,birk]{Martino Marisaldi}
\author[upc]{Oscar A. van der Velde}
\author[inaf]{Fabio Fuschino}

\cortext[cor1]{Corresponding author}

\address[upc]{Electrical Engineering Department, Universitat Polit\`ecnica de Catalunya, Barcelona, Spain.}
\address[inaf]{INAF-IASF, National Institute for Astrophysics, Bologna, Italy}
\address[birk]{Birkeland Centre for Space Science, University of Bergen,
Bergen, Norway}

%----------------------------------------------------------------------------------------
%	ABSTRACT
%----------------------------------------------------------------------------------------

\begin{abstract}
Global distribution of the Terrestrial Gamma-ray Flashes (TGFs) detected by AGILE and RHESSI for the period from March 2009 to July 2012 has been analysed. A fourth TGF production region has been distinguished over the Pacific. It is confirmed that TGF occurrence follows the Intertropical Convergence Zone(ITCZ) seasonal migration and prefers afternoons. TGF/lightning ratio appears to be lower over America than other regions suggesting that meteorological regional differences are important for the TGF production. Diurnal cycle of TGFs peaks in the afternoon suggesting that Convective Available Potential Energy (CAPE) and convection are important for TGF production. Moreover all AGILE passages over South America in the same period have been analysed to find meteorological preferences for TGF occurrence. In each passage the analysis of Cloud Top Altitude (CTA), CAPE, number of strokes, number of storms and coverage area of clouds with temperatures below \(-70^{\circ}\)C (Cloud Top Coverage area, CTC) are computed. On average, AGILE has been exposed to 19100 strokes between each TGF representing \(\sim\) 68 hours of exposure over active storms. High CAPE values, high cloud tops and high stroke occurrence suggests that meteorological conditions conducive to vigorous and electrically active storms are important for TGF production. It is shown that a high number of thunderstorms is preferable for TGF production which may be explained by a greater chance of the presence of a thunderstorm in the best development stage for TGF production. High tropopause altitude seems to be important but not primordial for TGF production.

\end{abstract}
\end{frontmatter}

%----------------------------------------------------------------------------------------
%	PAPER
%----------------------------------------------------------------------------------------

\section{Introduction}
CTR Wilson predicted in 1925 that electrons accelerated in high electric fields in thunderstorms would emit high energy radiation \citep{Wilson1925}. Many years later, this phenomena was observed by \citet{Fishman1994a}, who reported 12 high energy emissions coming from Earth detected by the Burst and Transient Source Experiment (BATSE) on the Compton Gamma Ray Observatory (CGRO), with durations of few milliseconds and one or multiple peaks. This phenomena, known nowadays as Terrestrial Gamma-Ray Flashes (TGFs), was first related to thunderstorms \citep{Fishman1994a}. Then, TGFs were related to sferics, electromagnetic lightning emissions  and sprites \citep{Inan1996a}. Later, correlations of TGFs with positive polarity lightning \citep{Cummer2005c} and directly to positive intra-cloud (IC) flashes with peak currents between \( 15 \, kA \) and \( 57 \, kA \) \citep{Stanley2006c}, refuted the theory proposed by \citet{Inan2005d} that TGFs were produced by electromagnetic pulses radiated by high return strokes with high peak currents (\(450-700\,kA\)). 
On the other hand, high energy photons detected from cloud-to-ground (CG) stepped leaders \citep{Moore2001,Dwyer2003f} and laboratory discharges \citep{Dwyer2005,Dwyer2008,March2010,March2011} suggest that TGFs may be produced by similar processes. Analysis of the TGFs spectra had restricted the production altitude to 15-20 km \citep{Dwyer2005c,Carlson2007c,Østgaard2008,Hazelton2009,Gjesteland2010}. Moreover \citet{Lu2010} and \citet{Shao2010} find evidences of TGFs occurring during first stages of IC flashes. These results leads to conclude that Bremsstrahlung radiation emitted by runaway  electrons during IC flashes is the most probable source of TGFs. The most accepted theories nowadays are the relativistic feedback mechanism \citep{Dwyer2003f,Dwyer2008a,Dwyer2012g} and lightning leader emission models \citep{Dwyer2008a,Dwyer2010c,Carlson2009e,Carlson2010f,Celestin2011}.
\\

Other satellite instruments, mainly designed for the study of high energy astrophysics, have detected TGFs. These new instruments have allowed to detect more events and better define their spectra and geographical occurrence. \citet{Smith2005} reported 86 TGFs detected by the Reuven Ramaty High Energy Spectroscopic Imager (RHESSI) \citep{Smith2002} with energies up to 20 MeV. These TGFs showed that the geographical occurrence is greater over tropical continental regions in good agreement with the global lightning occurrence. \citet{Smith2010} estimated a global daily occurrence on Earth of 50 TGFs. \citet{Gjesteland2012c} showed by an improvement of the search algorithm criteria for RHESSI measurements that the number of TGFs detected in the period 2004-2006 is twice the number previously reported. In 2006, \citet{Williams2006} ensured that IC flashes are the only candidates to produce TGFs because gamma rays produced at low altitudes by CG flashes would not be able to escape due to the atmospheric attenuation. This theory agrees with the higher TGF occurrence in the tropics where deep convection and high tropopause altitude allow high altitude thunderstorms and consequently occurrence of IC flashes higher in the atmosphere. On the other hand, \citet{Smith2010} showed using Monte Carlo simulations of gamma rays through atmosphere that tropopause altitude do not explain all the differences in the TGF geographical occurrence. Some other regional meteorological features may play an important role on these differences. Some parameters were studied by \citet{Splitt2010}. He reported 805 TGFs observed by RHESSI with geographical occurrence in very good agreement with the three lightning continental tropical chimneys \citep{Williams2005}. Moreover, diurnal cycle and seasonal latitudinal distribution of TGFs and lightning are very similar in the tropics. Convective Available Potential Energy (CAPE) and outgoing infra-red flux comparison of the 805 TGF sub-satellite locations and 805 tropical random locations at the times of TGFs showed that TGFs occur in higher CAPE values than tropical background. \citet{Splitt2010} also reported the top altitudes and areal extent of 29 thunderstorms producing TGFs. The altitude varied between 13.6 and 17.3 km. The areal varied between 400 km\(^2\) and 111100 km\(^2\) what showed that thunderstorms producing TGFs are not limited to larger convective systems. Finally a single case study of a thunderstorm producing a TGF was presented that showed an elevated mixed phase around 6 km.\\

The Mini-Calorimeter(MCAL) \citep{Labanti2009} on-board of the Astrorivelatore Gamma ad Imagini Leggero (AGILE) satellite  of the Italian Space Agency (ASI) \citep{Tavani2009} was launched on 23 April 2007 . First measurements reported a geographical distribution and diurnal cycle of TGFs consistent with RHESSI observations, and TGF photons with energies up to 40 MeV \citep{Marisaldi2010}. In 2011, \citet{Fuschino2011} demonstrated that TGF/lightning production rate was different in the three lightning chimneys, being greater over South America suggesting again that meteorological geographical differences play an important role in the TGF production. The most exceptional result of AGILE is its detection of photons up to 100 MeV \citep{Tavani2011}, whose spectrum is not consistent with Relativisitic Runaway Electron Avalanche (RREA) \citep{Gurevich1992,Dwyer2005c}. In 2014 the first official AGILE TGF catalog \citep{Marisaldi2014} was published.\\

In this paper we present two different analysis. The first is a global occurrence comparison between RHESSI and AGILE TGFs of diurnal cycle and monthly distribution. The main objective of this analysis is to find regional differences in TGF production which may be explained by other meteorological factors than lightning altitude production. Previous publications suggested that lightning altitude production is important in TGF production but other meteorological factor may play an important role \citep{Smith2010,Splitt2010,Fuschino2011}. To enhance the study about meteorological geographical differences that may play an important role in TGF production, a second analysis has been proposed. In this analysis different parameters that characterize atmospheric thunderstorm conditions are considered for all AGILE passages over South America during the period 2009-2012. The selected variables for the study are the Cloud Top Altitude (CTA), CAPE, number of strokes, number of storms and Coverage Area of clouds with temperatures below \(-70^{\circ}\)C (Cloud Top Coverage,CTC). The objective is to identify most prevalent meteorological characteristics conducive to TGF production. The main purpose of this study is compare meteorological characteristics during passes with TGFs against those in passes in which TGFs were absent.

\section{Data and Methodology}
\subsection{Global Analysis}
\subsubsection{TGF data}
For the first analysis, geographical distribution, monthly rate distribution and diurnal cycle of RHESSI and AGILE TGFs in the period of March 2009 to July 2012 have been compared, which is the period corresponding to the official AGILE catalog \citep{Marisaldi2014}. RHESSI is a National Aeronautics and Space Administration (NASA) spacecraft orbiting Earth in a low-Earth orbit (600 km) with an inclination of \(38^{\circ}\) launched 5 February 2002 designed to detect solar X and gamma rays. The instrument consists of nine Germanium detectors that can detect each photon in the range from 3 keV to 17 MeV individually without the need of an on-board trigger \citep{Smith2002}. In this paper we use the on-line catalogue of RHESSI TGFs provided by David M. Smith (http://sprg.ssl.berkeley.edu/~dsmith), with 283 TGFs in the period of March 2009 to July 2012.

AGILE is a space mission of the Italian Space Agency with four different instruments to study gamma ray astrophysics. It was launched on 23 April 2007 in a low-Earth orbit at 550 km altitude with an inclination of \(2.5^{\circ}\), therefore covering with high exposure a very narrow band across the equator. The main instrument used for TGF detection is the MCAL \citep{Labanti2009} and is formed by 30 CsI(Tl) scintillator bars with an energy range from 300 keV to 100 MeV. We use the official AGILE TGF catalog \citep{Marisaldi2014}, 308 TGFs, with maximum photon energy 30 MeV, detected in the period of March 2009 to July 2012, available at http://www.asdc.asi.it/mcaltgfcat/. 

\subsubsection{Lightning Data}

For the same period, lightning data of the Very Low Frequency lightning detection network called World Wide Lightning Location Network (WWLLN, \citep{Rodger2006}) is used. In order to be compared with RHESSI and AGILE distributions WWLLN strokes have been analyzed in latitudinal ranges covered by both satellites, \(\pm 44^{\circ}\) for RHESSI and \(\pm 8^{\circ}\) for AGILE, the extra latitude interval with respect to satellite orbital inclination accounts for a maximum TGF satellite footprint distance in the order of 600 km. A relative WWLLN efficiency is provided by \citet{Hutchins2012} and used for the global longitudinal distribution. This detection efficiency is calculated from 14 April 2009 up to know. Therefore, the global longitudinal distribution plot presents RHESSI, AGILE and WWLLN from 14 April 2009 to 31 July 2012. For the latitudinal and diurnal cycle graphics data is plotted from March 2009 to July 2012. Finally, the monthly rate distribution for three years of data, from March 2009 to February 2012, has been analyzed.

\subsection{AGILE passages}
\subsubsection{AGILE passages and GOES-East images}
\begin{figure}
 \noindent\includegraphics[width=1\linewidth]{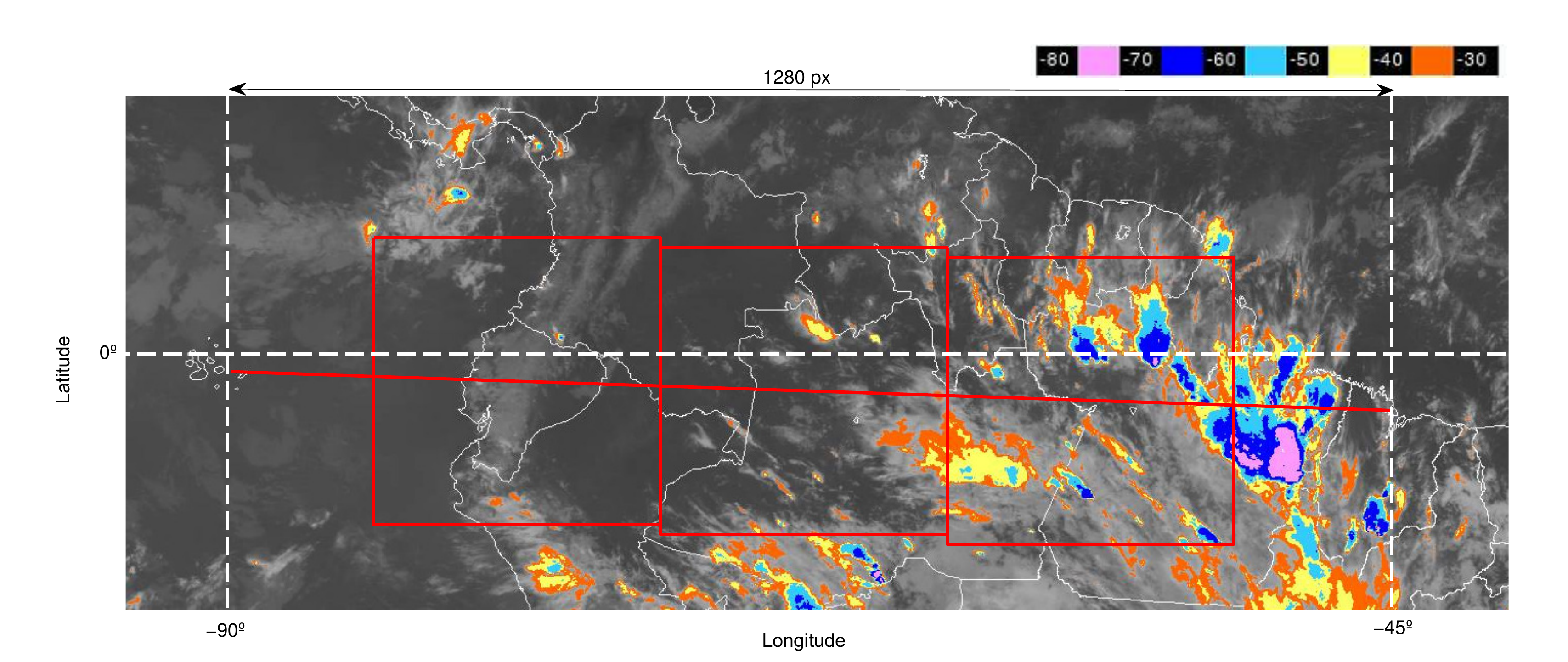}
 \caption{Example of passage analysed. Red line is the satellite orbit and red squares are the squares of 280 pixels side (1230 km) used in the analysis.}
 \label{fig:meth}
 \end{figure}

For the period from from March 2009 to July 2012 a total of 15951 passages over South America are considered. The data provided is a set of sub-satellite geographic coordinates and UTC time where \(t_1\) corresponds to the satellite passage at \(-90^{\circ}\) longitude and \(t_2\) corresponds to \(-45^{\circ}\) longitude for each passage. Infra-red images of channel 4 of the GOES-East satellite have been also used. GOES-East is a geosynchronous satellite at \(0^{\circ}\) latitude and \(-75^{\circ}\) longitude at 35800 km altitude that scans half the Earth every 15 minutes. For each AGILE pass over South America the corresponding GOES images was chosen. Taking four geographic reference points a function was used to transform passage geographic coordinates to GOES image pixels. Each AGILE passage is then \(\sim \) 1120 pixels wide in longitude. These 1120 pixels corresponds to \(-45^{\circ}\) longitude \(\sim\) 4995 km (\(1^{\circ}\) equals 111 km at the equator). That means that the GOES images used here have a pixel resolution of 4.4 x 4.4 km\(^2\). The analysis of all the parameters is implemented in three squares of 1230 km side equidistant to \(-90^{\circ}\) and \(-45^{\circ}\) longitude. 1230 km side is the maximum distance of a point on the Earth surface from the AGILE subsatellite point that can be seen at the satellite altitude, hereafter called field of view (FoV) for the purposes of the paper (figure \ref{fig:meth}). These squares are called AGILE FoV squares (figure \ref{fig:meth}). Knowing \(t_1\) and \(t_2\) and approximating AGILE track as a straight line, which is a reasonable assumption given the low orbital inclination, it is possible to calculate the time \(t_{i} \) (\(i=1,2,3\)) where AGILE satellite is in the center of each square: \(t_{i}=t_2+(t_2-t_1)/4 i\), the total time \(t_t\) required by the satellite to cover the square, \(t_t=(t_2-t_1)/4\) and the times \(t_{en_i} \) and \(t_{ex_i}\) that satellite enters and exits each square, \(t_{en_i}=t_i-t_t/2\) and \(t_{ex_i}=t_i+t_c/2\). The analysis of each variable is then performed inside each AGILE FoV square.
There are a total of 47853 AGILE FoV squares without a TGF  detected. For the 44 passages with a TGF detected it is taken only one AGILE FoV square. This one is centered at the substallite position when AGILE detected the TGF.   
  
\subsubsection{Cloud Top Altitude and Coverage}
  
Cloud analysis is done with GOES-East images of infra-red channel 4 (\(10.20 \, \mu m\) - (\(11.20 \,  \mu m\)) provided by the Instituto Nacional de Pesquisas Espaciais (INPE) of Brazil (http://satelite.cptec.inpe.br). These images provide cloud top temperatures coloured in a range from \(-30^{\circ}\)C to \(-80^{\circ}\)C with \(10^{\circ}\)C intervals (figure \ref{fig:meth}). For each square the area covered by cloud tops with temperatures below \(-70^{\circ}\)C is calculated by simply counting the pixels corresponding to these temperatures and multiplying by pixel area that is 4.4 x 4.4 km\(^2\). If this area is larger than 0 it means that at least one high topped thunderstorm exists below the AGILE satellite. This calculation gives an approximation of the total area covered by high-topped thunderstorms.\\

Because radiosonde stations are scarce in tropical regions, we consider models and reanalysis. MSISE-90 model is employed to infer Cloud Top Altitudes from temperatures given by GOES satellite infrared images \linebreak (http://ccmc.gsfc.nasa.gov/modelweb/atmos/msise.html). For each square it is assumed that the atmosphere is in equilibrium. A vertical profile of neutral temperature is given by the MSISE-90 model at the center of the square at the time the satellite is passing over it, obtaining altitudes corresponding to GOES satellite infra-red image temperatures, that allows to identify maximum Cloud Top Altitudes in the square of thunderstorms when a TGF is detected or not.\\

There were only GOES-East images available for 8577 AGILE passages. Cloud Top Altitude and Coverage have been analysed for 25731 AGILE FoV squares without TGF and 23 with TGF. For the analysis of the number of strokes, number of thunderstorms and CAPE, GOES-East data is not needed.

\subsubsection{Strokes and Thunderstorms}
Data from WWLLN is used to check lightning and thunderstorms. The total number of lightning strokes in the FoV of AGILE satellite when is over each square is the number of detections by WWLLN inside the square between times \(t_{en_i}\) and \(t_{ex_i}\). The mean value of \(t_{ex_i} - t_{en_i}\) is 3.2 minutes. To calculate the number of storms on each square, the total number of strokes inside each square in the period \(t_{en_i} - 20 \) minutes and \(t_{ex_i}\ + 20\) minutes has been found, clustering these strokes using a commercial software package (MATLAB R2012b, The MathWorks Inc., 2012) own function that calculates centroid distance of the cluster (Unweighted Pair Groups with Arithmetic Mean). The criteria values to build the storms are a maximum centroid distance of \(1^{\circ}\) and at least 5 strokes.  
   
\subsubsection{CAPE}   
   
CAPE data is obtained from European Centre for Medium-Range Weather Forecasts (ECMWF, http://www.ecmwf.int/research/era/do/get/era-interim), defined as "amount of potential energy an air parcels acquires when lifted adiabatically from its lifting condensation level to the level of neutral buoyancy". This data is given every 3 hours with \(0.7^{\circ}\) x \(0.7^{\circ}\) resolution. We take the maximum CAPE before the satellite passage inside each square. This CAPE value will be between 3h to 0h before the satellite passage. Although it can be far from a TGF detection, previous CAPE better describes energetic conditions prior to storms producing TGF because CAPE is removes as storms restabilize the environment. 

\section{Results and Discussion}
\subsection{Global Analysis}
\begin{figure}
 \noindent\includegraphics[width=1\linewidth]{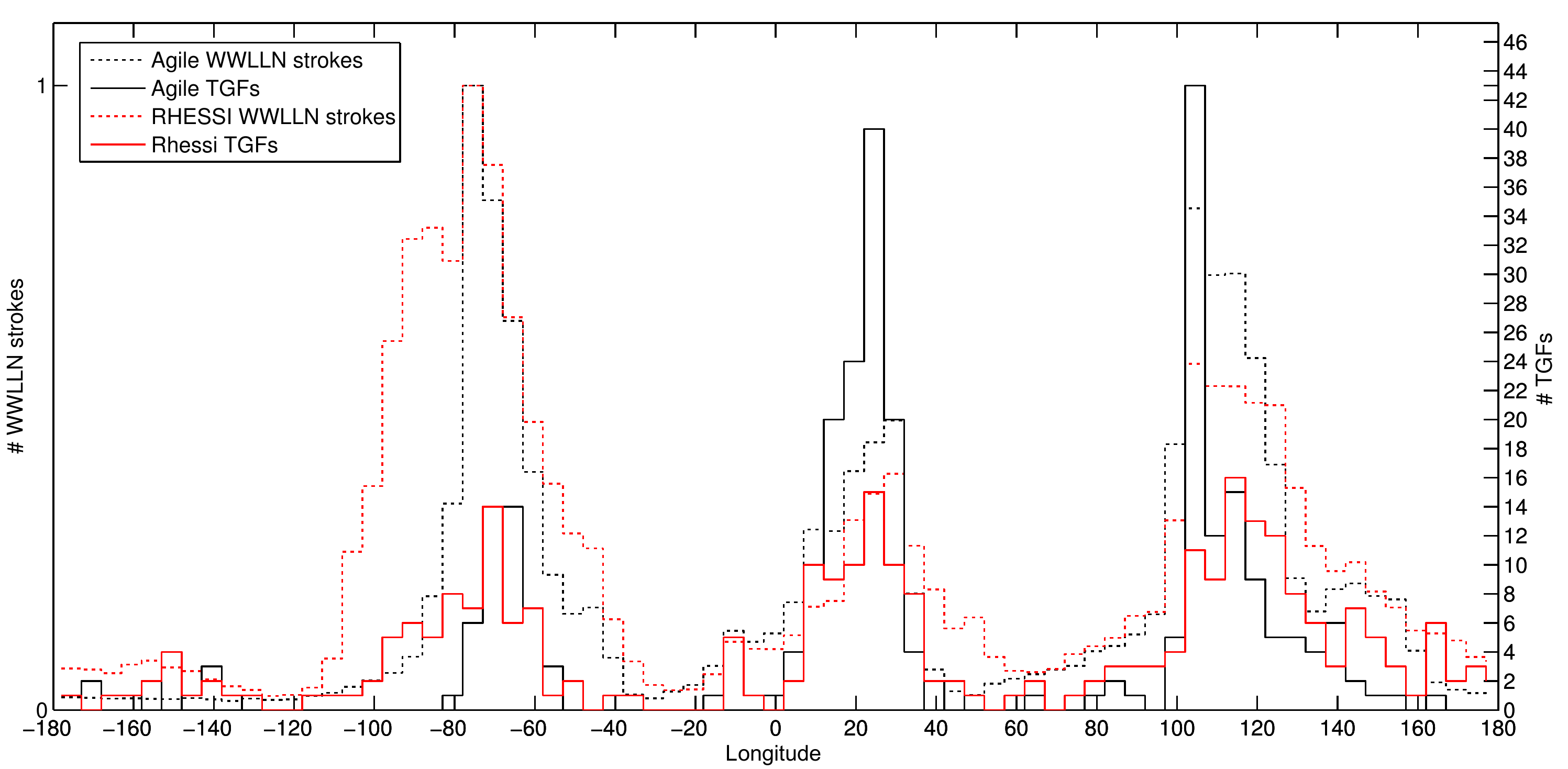}
 \caption{Longitudinal distribution for April 2009 - July 2012 period of AGILE WWLLN strokes (black dashed line), AGILE TGFs (black line), RHESSI WWLLN strokes (red dashed line) and RHESSI TGFs (red line).}
 \label{fig:gen}
 \end{figure}

\begin{figure}
 \noindent\includegraphics[width=1\linewidth]{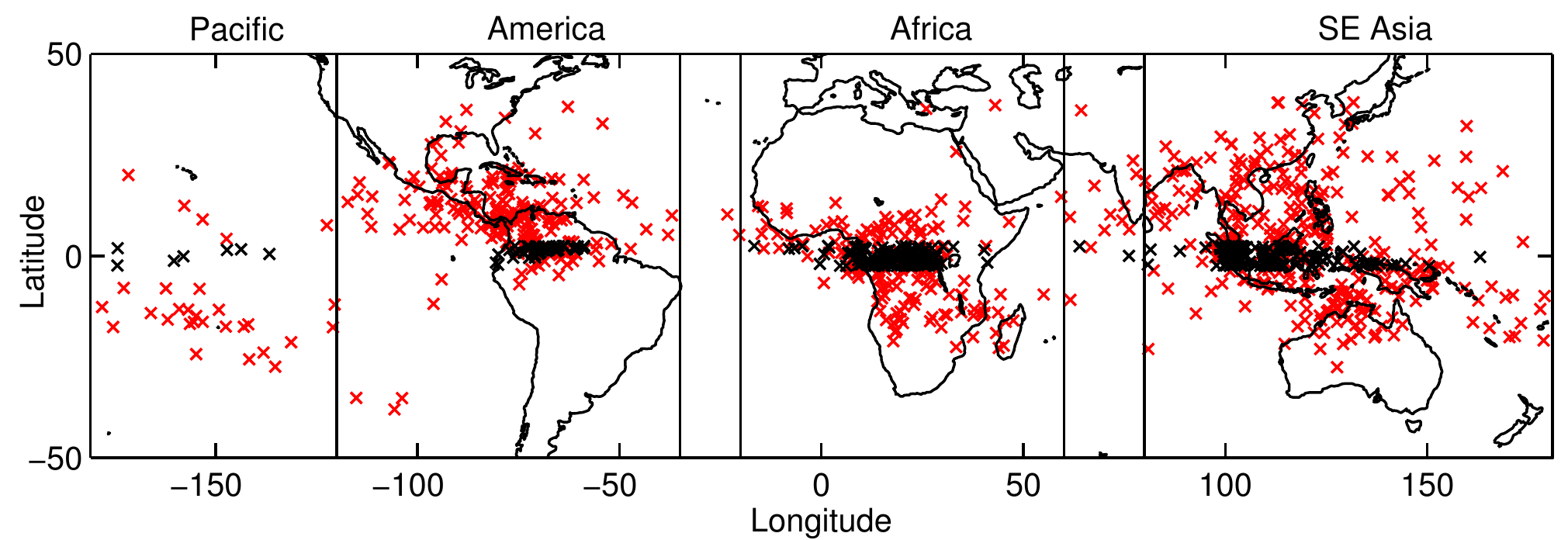}
 \caption{Geographical distribution for March 2009 - July 2012 period of RHESSI TGFs (red crosses) and AGILE TGFs (black crosses). Vertical black lines limits latitudes used for regional analysis}
 \label{fig:latlon}
 \end{figure}

Geographic longitudinal distribution of 294 and 283 TGFs detected by AGILE and RHESSI respectively, and WWLLN strokes for the April 2009 - July 2012 period are plotted in figure  \ref{fig:gen}. The analysis takes into account the differences in detection efficiency of the network in different regions, depending on the sensors geographical density, as well as the variation of efficiency at different times, when new sensors are added to the network. The relative detection efficiency of WWLLN in the whole Earth is provided by WWLLN team \citep{Hutchins2012}. This relative detection efficiency for WWLLN is calculated on a 1 degree grid for each hour. This data has been used to normalize the stroke density in figure \ref{fig:gen}. It is easy to identify in figure \ref{fig:latlon} three TGF producing regions  which are the tropical continental regions America, Africa and Asia. Furthermore, a fourth TGF region is identified over the Pacific. This region appeared previously in \citet{Collier2011} but it was not mentioned. Although this region is less active on TGF production than the other three, TGF and WWLLN distributions shows relevant maxima compared to other oceanic regions. TGF to lightning ratio is clearly lower over America compared to the other two continental regions for both AGILE and RHESSI. This result differs from \citet{Fuschino2011} (AGILE) and \citet{Smith2010} (RHESSI) who found a higher TGF/lightning ratio of TGFs over America compared to other regions. \citet{Fuschino2011} took only lightning data for AGILE inclination region, from \(-2.5^{\circ}\) to \(2.5^{\circ}\) latitude, unlike here where lightning data is considered from \(-8^{\circ}\) to \(8^{\circ}\) latitude, covering whole AGILE FoV. However, \citet{Fuschino2011} and \citet{Smith2010} compared data from the LIS (Lightning
Imaging Sensor) and OTD (Optical Transient Detector) instruments that detects total lightning (CG and IC) while here we are using WWLLN data that mainly corresponds to highly energetic CG flashes.\\
It can also be observed in figure \ref{fig:gen} that the four AGILE production regions have single narrow peaks unlike RHESSI, which has wider distributions with several peaks. This difference is due to the difference in satellite inclinations (figure \ref{fig:latlon}). Becasue RHESSI scans a wider range of latitudes there is land at more longitudes while lightning and TGF have higher occurrence over land and coastal regions than far out over the sea. Figure \ref{fig:latlon} also shows longitudinal limits used in the range of latitudes for RHESSI analysis of latitudinal distribution, monthly distribution and diurnal cycle for the four regions separately. These limits correspond to the minima of distributions in figure \ref{fig:gen}. These limits are from \(-180^{\circ}\) to \(-120^{\circ}\) for the Pacific, from \(-120^{\circ}\) to \(-35^{\circ}\) for America, from \(-20^{\circ}\) to \(60^{\circ}\) for Africa and from \(80^{\circ}\) to \(180^{\circ}\) for Asia.\\
   
\begin{figure}
 \noindent\includegraphics[width=1\linewidth]{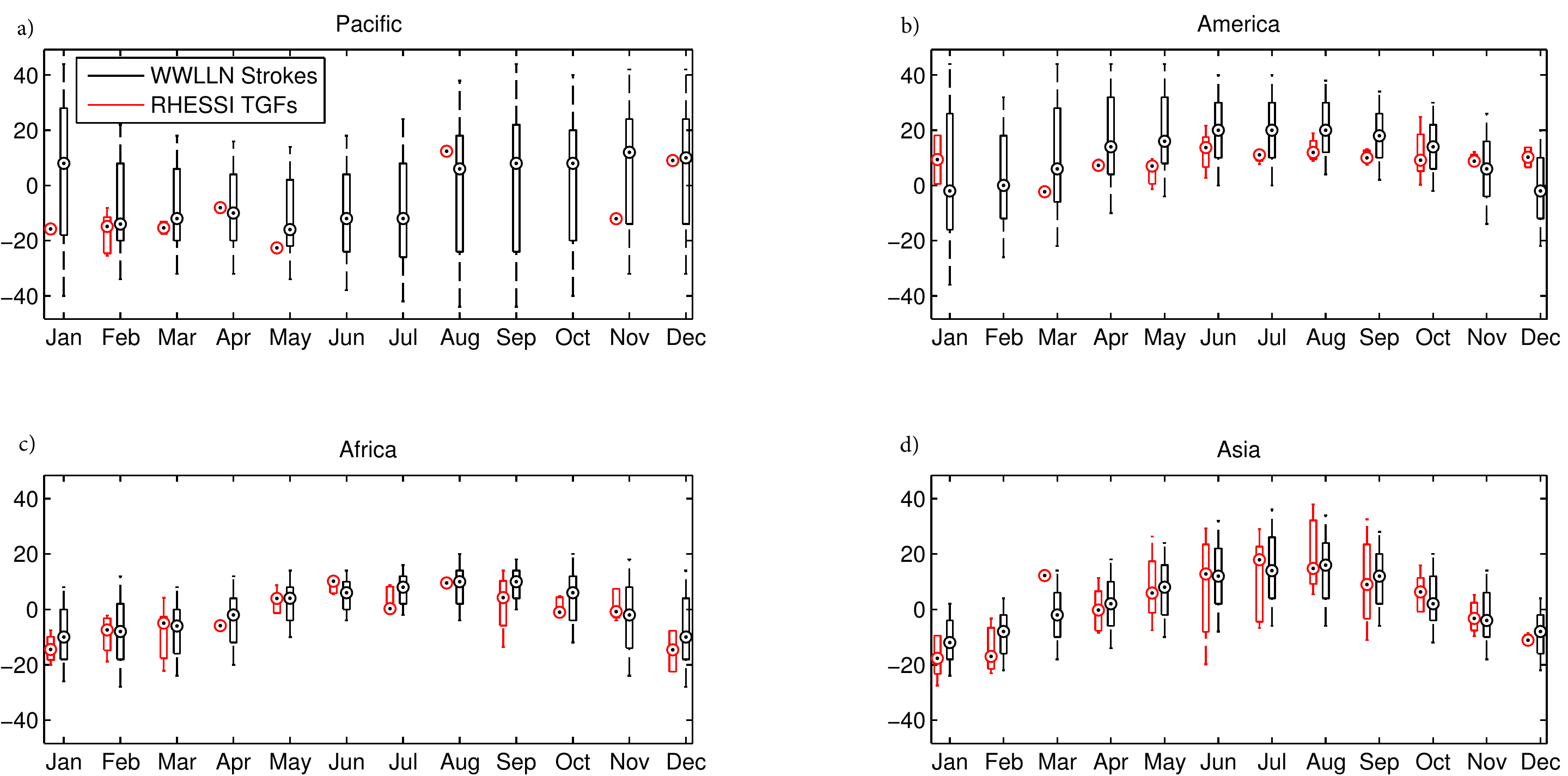}
 \caption{Monthly boxplot graphic for March 2009 - July 2012 period of latitudinal distribution for WWLLN strokes (black) and RHESSI TGFs (red). Circle is the median, box is the 50\% of the data and whiskers around 95\% of the data.}
 \label{fig:box}
 \end{figure}

Figure \ref{fig:box} shows the monthly latitudinal distribution within the four TGF production regions in a box-plot graph for the 2009 - 2012 period (note that the box-plot graph does not take into account the total number TGFs, only median and dispersion over latitude). This box-plot graph is presented only for RHESSI TGFs, since AGILE inclination is too small for this kind of analysis. TGF detections follow the same latitudinal migration throughout the year as the ITCZ \citep{Asnani1993}, being in June-August in the north and in December-February  in the south. This result was already reported by \citet{Splitt2010}. Figure \ref{fig:box} also shows the monthly latitudinal spread of lightning production, according to WWLLN data, in the four regions as well as the TGF production deficiency in North America reported by \citet{Smith2010}. For the Pacific region, TGFs occur at a wide range of latitudes consistent  with a larger spread in lightning detections. In America, when ITCZ is in the south, TGFs are detected in a narrower range of latitudes probably because RHESSI is affected by South Atlantic Anomaly (SAA). On the contrary, when the ITCZ is in the north  TGFs are detected in a wider range of latitudes although there are few in North America. African TGF and lightning  production is clearly limited north of the equator by the existence of the Sahara desert. In the SE Pacific region, a wide range of latitudes can be observed in all months, although for latitudes above \(20^{\circ}\) or below \(-20^{\circ}\) there are less TGFs than expected in relation to lightning production. This could be associated with the lowering of tropopause height at higher latitudes \citep{Smith2010}. In the four TGF production regions, medians of the locations do not follow a symmetric distribution with respect to the equator, but are displaced to the north, the same as the ITCZ.

\begin{figure}
 \noindent\includegraphics[width=1\linewidth]{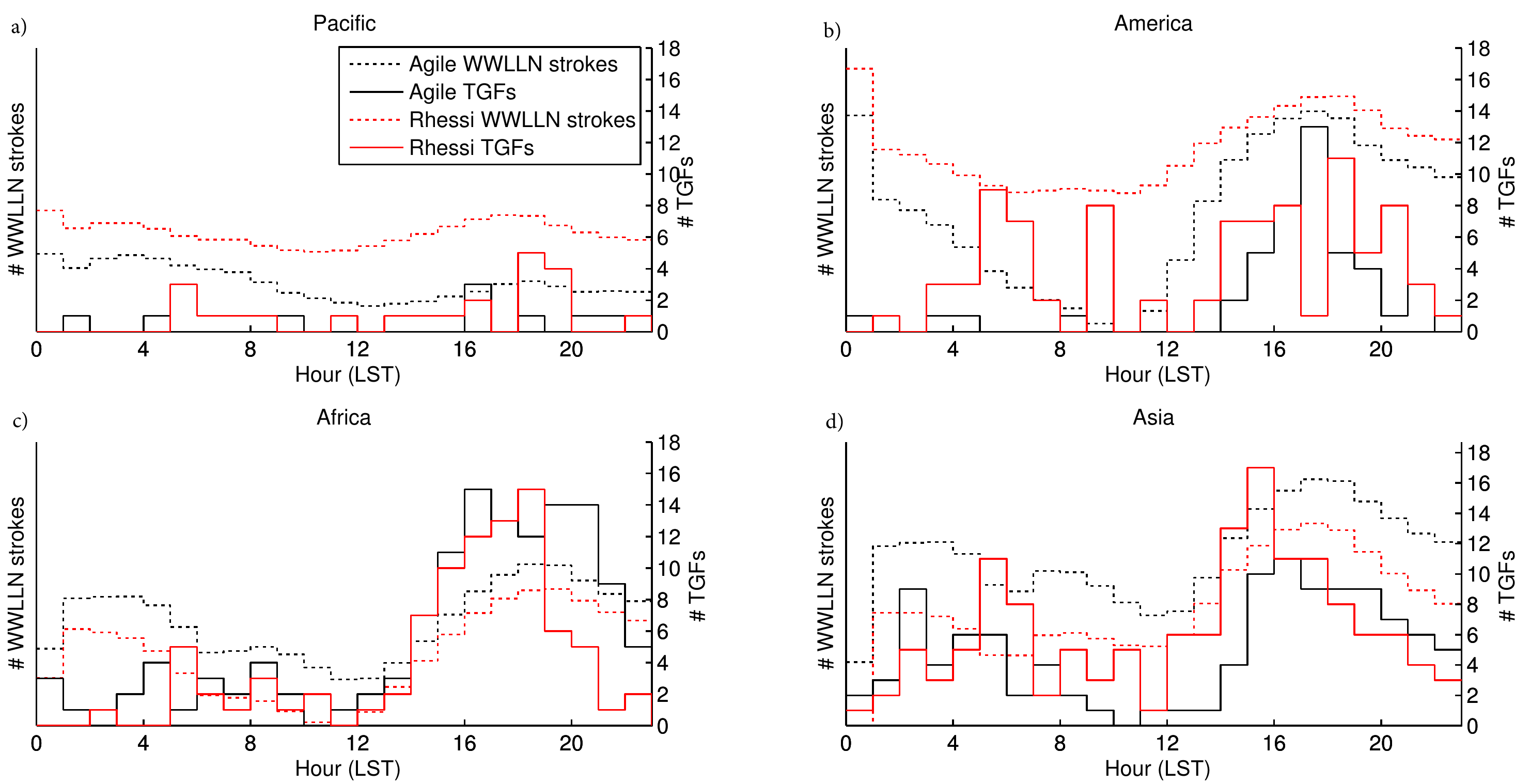}
 \caption{Diurnal cycle for March 2009 - July 2012 period  for the 4 TGF production regions of AGILE WWLLN strokes (black dashed line), AGILE TGFs (black line), RHESSI WWLLN strokes (red dashed line) and RHESSI TGFs (red line).}
 \label{fig:genhh}
 \end{figure}

In figure \ref{fig:genhh} the diurnal cycle for AGILE TGFs, RHESSI TGFs and WWLLN strokes in each region are plotted for the March 2009 - July 2012 period. In the four regions, AGILE, RHESSI and WWLLN stroke distributions agree with the Diurnal Tropical Cycle of Precipitation (DTCP) which is dominated by a maximum during the  afternoon due to thunderstorms and a secondary maximum during night or early morning \citep{Asnani1993}. It is complicated to analyse the Pacific region due to the low number of TGFs. In the other three regions the RHESSI TGF distribution has two peaks, the largest peak in the afternoon and the smaller peak in the early morning. In America the peaks are very clear. The maximum in the afternoon coincides with the maximum of WWLLN strokes while the maximum in the night coincides with minimum of WWLLN strokes distribution. The same behaviour can be observed for RHESSI and WWLLN strokes distributions in figure \ref{fig:genhh} in Africa and America. \cite{Splitt2010} already reported the coincidence of this secondary peak of TGFs in the early morning with low lightning activity. The AGILE TGF distributions also show the predominant peak of TGFs during the night within the three subregions, which coincide with the predominant peak of WWLLN stroke distributions. In South America veery AGILE TGFs were observed during the night and early morning. Finally, in Africa there is an observable peak of AGILE TGFs in the night but is not coincident  with the secondary maximum of WWLLN strokes at 2 LST. As will be seen in the following section, TGF production seems to be directly affected by the presence of large CAPE which in turn tends to be maximized the afternoons as a result of the solar heating.

The peaks March-May and September-November in the figure \ref{fig:genmm} correspond to the ITCZ crossing the equatorial region covered by AGILE. RHESSI distributions are more flat due to the larger inclination of the satellite which in turn makes it less influenced by the ITCZ movement. However over America during December to February TGF production almost drops to zero. During this season the ITCZ is south of the equator where TGF production should have been higher but satellite detection capabilities are affected by the South Atlantic Anomaly (SAA).

\subsection{AGILE passages analysis}

In this section the results for the meteorological conditions conducive to TGF production over South America are presented. The adopted method is the analysis and comparison of the meteorological conditions corresponding to AGILE passages related to TGF detections and those with not. Therefore there have been eliminated from the analysis all the FoV without TGF detections without storms. \( 61 \% \) of total FoV without a TGF detected have at least one thunderstorm.\\

44 TGFs have been observed for 838708 WWLLN strokes during a total exposure time of \(\sim 2550\) hours (this time is obtained by multiplying mean AGILE FoV time exposition, 3.2 minutes, by total number of AGILE FoV squares). This corresponds 1 TGF detected by AGILE for 19062 strokes observed by WWLLN during \(\sim 58\) hours of exposure. It is important to consider that global WWLLN efficiency is about \(\sim 10 \% \) \citep{Virts2013} which means that AGILE is detecting \(\sim\) 1 TGF every 191000 strokes and \(\sim 58\) hours of exposure. These ratio is remarkably lower than that reported in \citep{Fuschino2011} , based on LIS/OTD data and correlation studies. We argue that the ratio reported here is underestimated because strokes are not weighted for the different detection probability given by the different distance to the satellite footprint. Stroke occurrence (flash km \(^2\) yr \(^{-1}\)) is two orders of magnitude higher in equatorial South America than in Europe \citep{Abarca2010}. That means that same satellite over Europe would need \(\sim 5800 \) hours (\(\sim 242\) days) of exposure to observe a TGF, neglecting the further contribution of meteorological factors. This analysis suggests that a very large effective area is a crucial requirement for space instruments designed to detect TGFs above Europe.

\begin{figure}
 \noindent\includegraphics[width=1\linewidth]{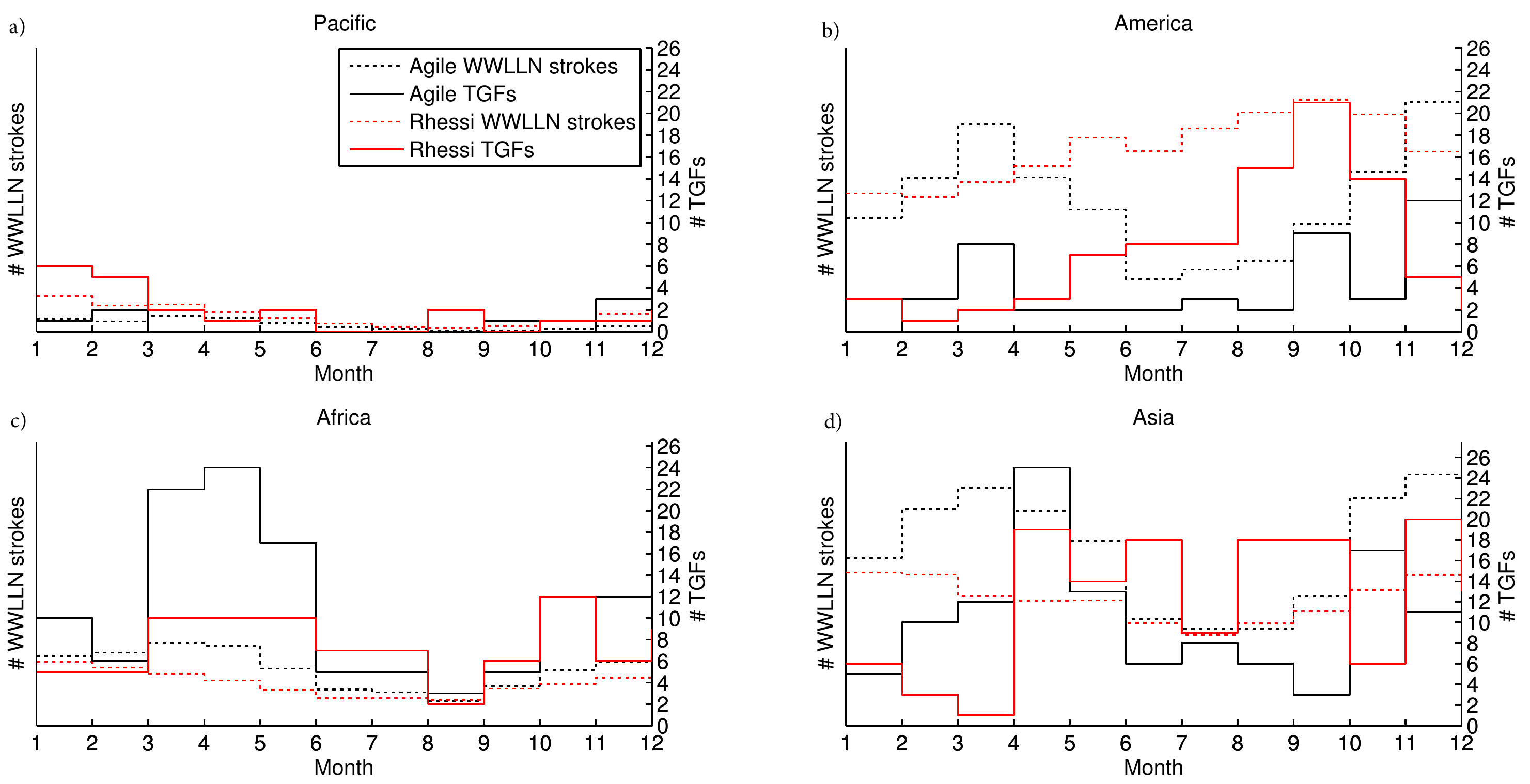}
 \caption{Monthly distribution for March 2009 - February 2012 period for the 4 TGF production regions of AGILE WWLLN strokes (black dashed line), AGILE TGFs (black line), RHESSI WWLLN strokes (red dashed line) and RHESSI TGFs (red line).}
 \label{fig:genmm}
 \end{figure}

\begin{figure}
 \noindent\includegraphics[width=1\linewidth]{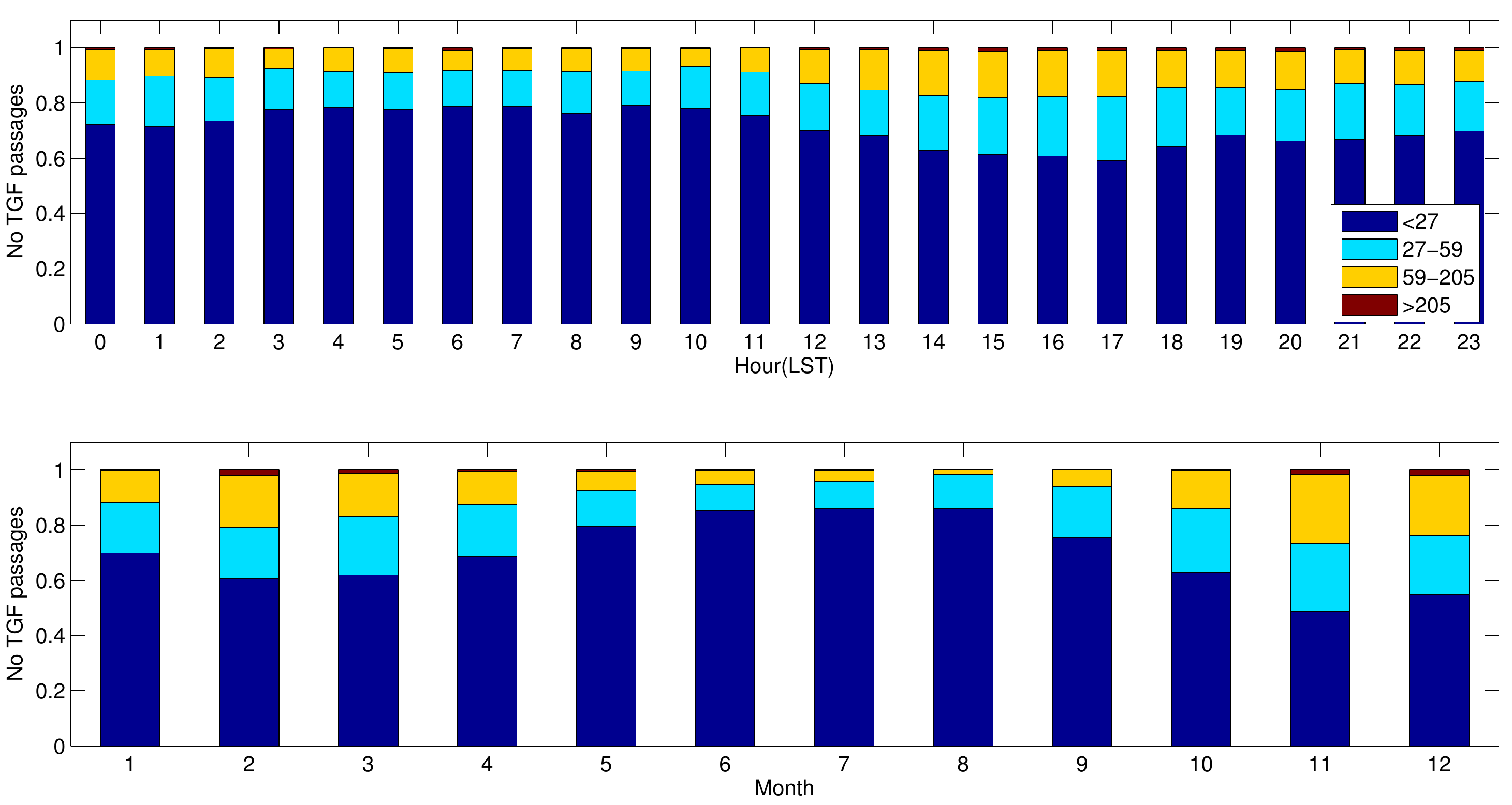}
 \caption{Number of strokes. Diurnal cycle (top) and monthly distribution (bottom) of the number of strokes of AGILE passages non detecting TGFs for March 2009 - July 2012 period over South America. Values of the 4 ranges correspond to AGILE passages detecting TGFs ( table \ref{tab:pctl})}
 \label{fig:str}
 \end{figure}

\begin{figure}
 \noindent\includegraphics[width=1\linewidth]{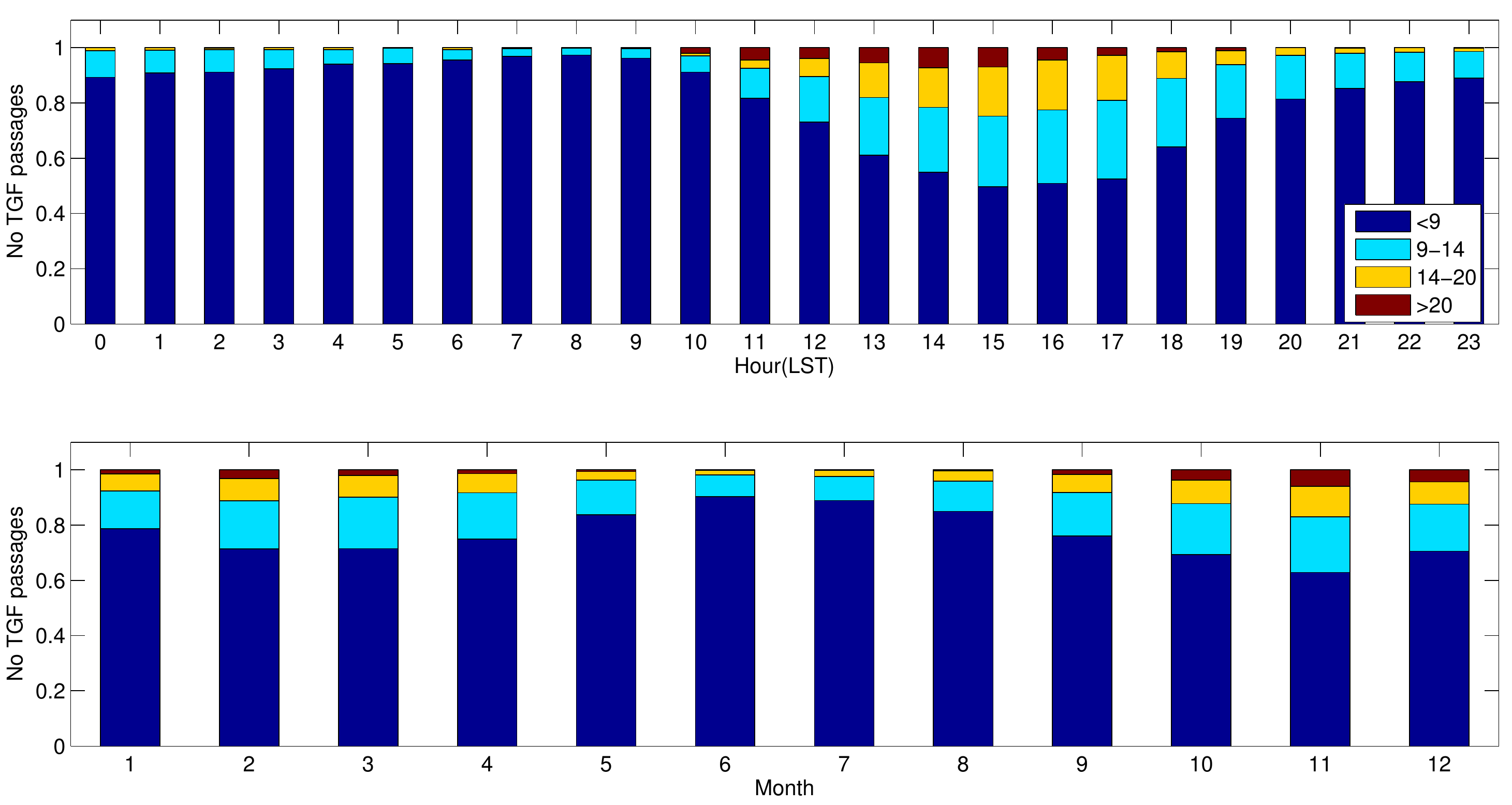}
 \caption{Number of storms. Diurnal cycle (top) and monthly distribution (bottom) of the number of storms of AGILE passages non detecting TGFs for March 2009 - July 2012 period over South America. Values of the 4 ranges correspond to AGILE passages detecting TGFs ( table \ref{tab:pctl})}
 \label{fig:sto}
 \end{figure}

\begin{figure}
 \noindent\includegraphics[width=1\linewidth]{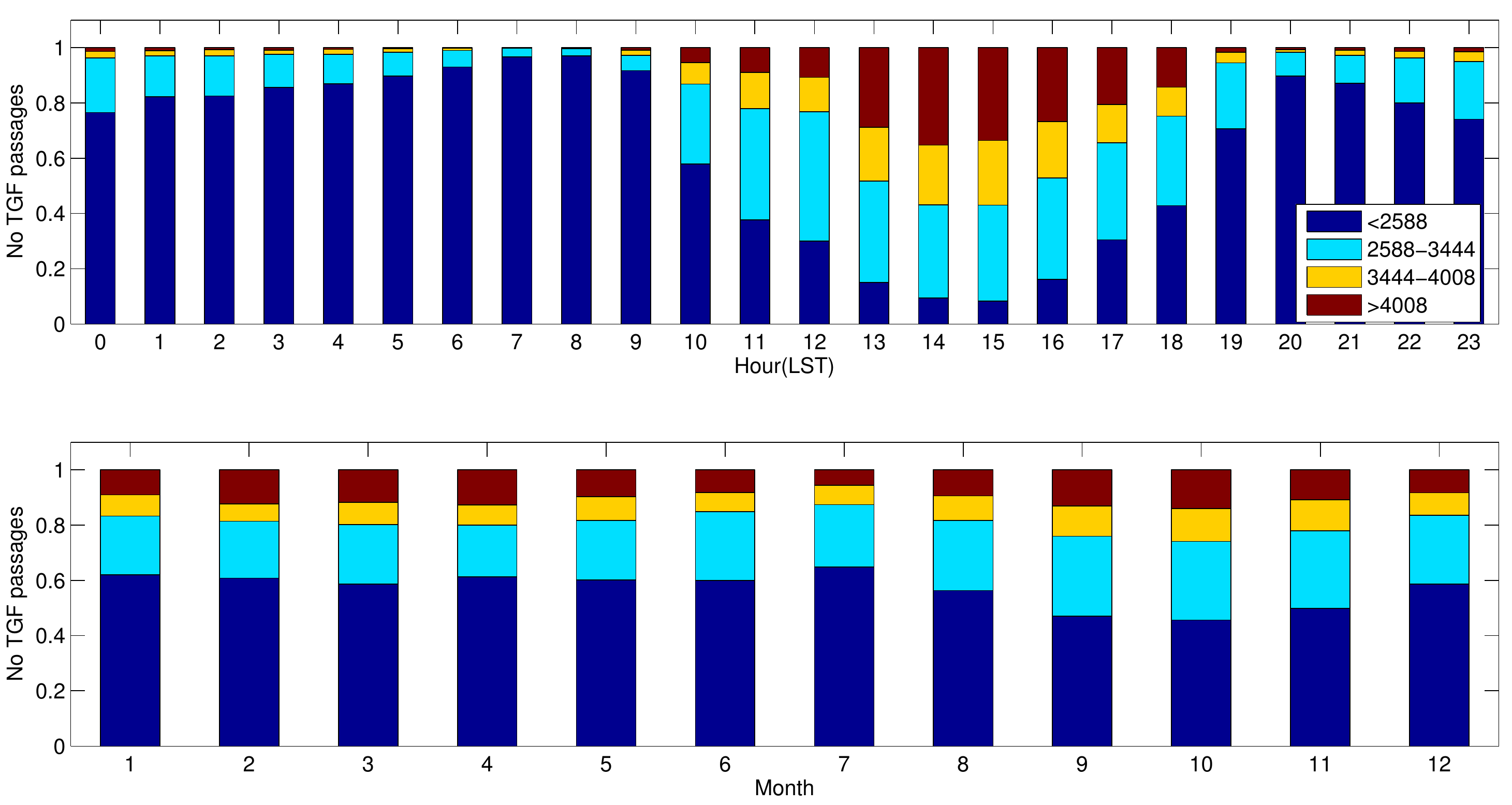}
 \caption{CAPE. Diurnal cycle (top) and monthly distribution (bottom) of the CAPE of AGILE passages non detecting TGFs for March 2009 - July 2012 period over South America. Values of the 4 ranges correspond to AGILE passages detecting TGFs ( table \ref{tab:pctl})}
 \label{fig:cape}
 \end{figure}

\begin{figure}
 \noindent\includegraphics[width=1\linewidth]{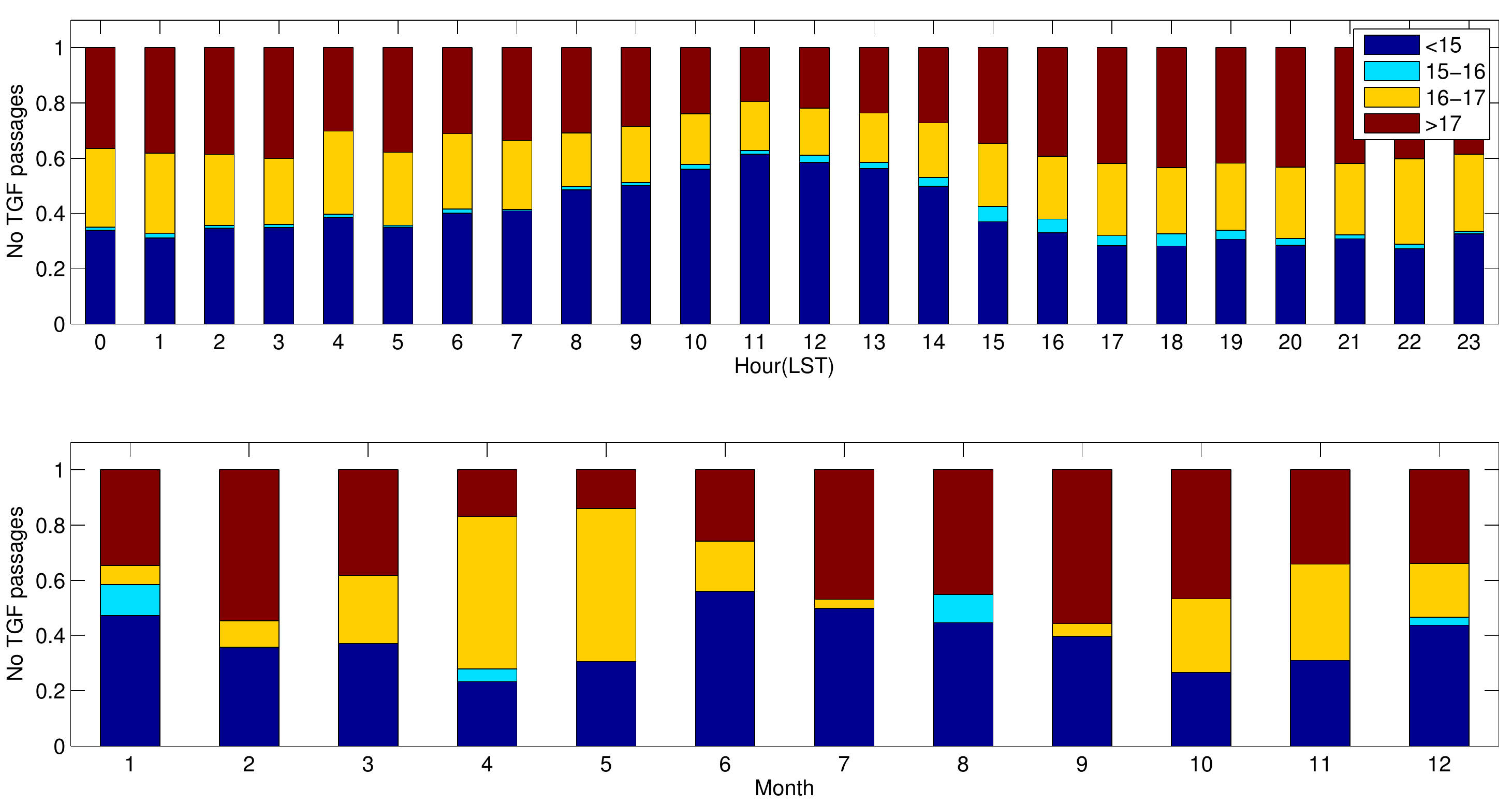}
 \caption{Cloud Top Altitude. Diurnal cycle (top) and monthly distribution (bottom) of the CTA of AGILE passages non detecting TGFs for March 2009 - July 2012 period over South America. Values of the 4 ranges correspond to AGILE passages detecting TGFs ( table \ref{tab:pctl})}
 \label{fig:alt}
 \end{figure}

\begin{figure}
 \noindent\includegraphics[width=1\linewidth]{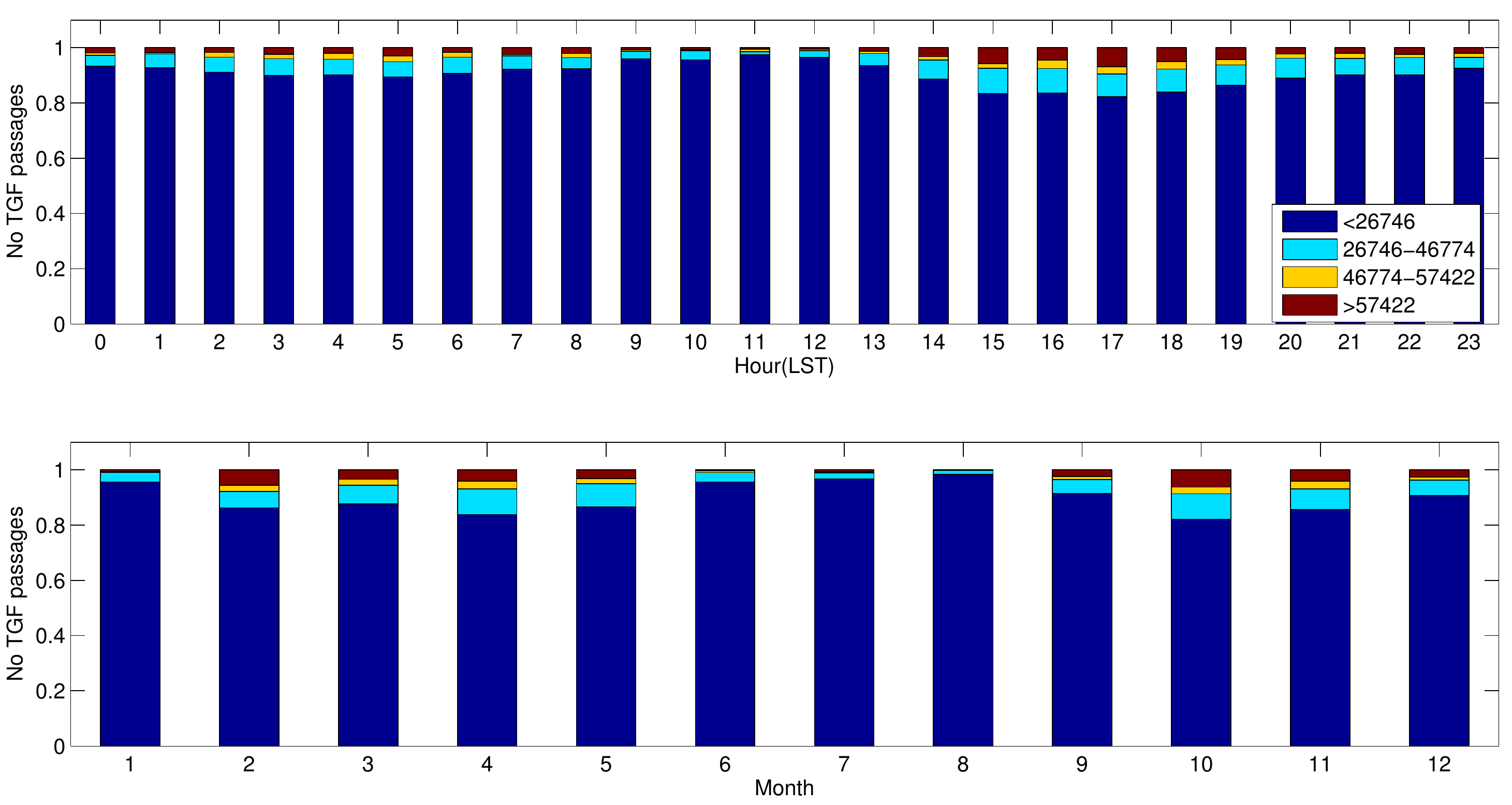}
 \caption{Cloud Top Coverage. Diurnal cycle (top) and monthly distribution (bottom) of the CTC of AGILE passages non detecting TGFs for March 2009 - July 2012 period over South America. Values of the 4 ranges correspond to AGILE passages detecting TGFs ( table \ref{tab:pctl})}
 \label{fig:area}
 \end{figure}

Table \ref{tab:pctl} shows the percentiles 10, 25, 50, 75 and 90 for the five meteorological parameters during AGILE passes with and without detected TGFs analysed: number of strokes, number of storms, Cloud Top Area (CTA), Covective Availabel Potential Energy(CAPE) and Cloud Top Coverage (CTC). Table \ref{tab:pctl2} shows percentiles of the same five meteorological parameters for the AGILE passages without detected TGFs corresponding to percentile values of AGILE passages detecting TGFs of table \ref{tab:pctl}. For example, according to table \ref{tab:pctl}, the \( 25 \% \) percentile for the number of strokes in the case of TGF corresponds to 27. This number corresponds to the \( 69 \% \) percentile in the case of passages without TGF detection, see table \ref{tab:pctl2}. Figures \ref{fig:str} - \ref{fig:area} show diurnal cycle and monthly distributions of the same meteorological parameters for passages non detecting TGFs. These distributions are plotted in stacked bar graphs of four categories. These categories are four ranges corresponding to quartiles and median values of detecting TGF passages. In this way we can figure out if diurnal cycle and monthly distribution of non detecting TGF passages agrees with detecting TGF distributions of figures \ref{fig:genhh} and \ref{fig:genmm}. Since TGFs have a clear diurnal and monthly signature (figures \ref{fig:genhh} and \ref{fig:genmm}), it is important to assess this behavior also for non-detecting TGF passages.  

\subsubsection{Number of Strokes and Storms}

\( 75 \% \) of TGFs were detected when the AGILE satellite was exposed to more than 80 strokes/minute (27 strokes in 3.2 minutes, time over AGILE FoV square, and WWLLN efficiency \(\sim 10 \% \) \citep{Virts2013}) and 9 active storms (table \ref{tab:pctl}).   This is because of the more strokes and thunderstorms the higher is the probability to detect a TGF. Table \ref{tab:pctl2} shows that these situations corresponding to \( 75 \% \) TGFs detected represents only less than \( 31 \% \) of AGILE passages non detecting TGFs, meaning that TGFs are detected when satellite is exposed to higher strokes and storms rates. Moreover, in figures \ref{fig:str} and \ref{fig:sto} it is shown that number of strokes and number of storms of AGILE passages non detecting TGFs corresponding to \( 75 \% \) values of TGFs detected, peaks clearly in the afternoon as TGFs detected do. For monthly distribution also exists coincidence for the same values as diurnal cycle, although peaks are less pronounced.

\subsubsection{CAPE}

Table \ref{tab:pctl} shows that \( 75 \% \) of TGFs detected occurred when CAPE was higher than 2588 J/kg. This range corresponds to \( 43 \% \) of AGILE passages without detected TGFs. Figure \ref{fig:cape} shows that passages with CAPE values higher than 2588 J/kg tend to occur mostly in the afternoon without a monthly preference. These results reinforce what it was shown in the previous section, i.e. that TGFs mostly occurs at afternoon when solar irradiance results in high CAPE. So solar irradiance and CAPE plays and important role in helping to get the right conditions in thunderstorms for the TGF production and could explain the differences of regional TGF/lightning ratio presented previously. It is also worth noting that the largest CAPE values are observed at 15 LST, with a distribution very similar to that corresponding to passages with TGF detection. However, the peak in TGF detection is two hours later, at 17 LST. This delay is possibly due to thunderstorm lightning activity evolution driven by high CAPE values. In fact, as shown in figure \ref{fig:genhh}, lightning activity over South America peaks at 17 LST.

\subsubsection{CTA}

\( 50 \% \) of TGFs occurred when CTA was \(>\)16 km (table \ref{tab:pctl}). On the other hand, CTA \(>\)16 km represents a \( 60 \% \) of total AGILE passages (table \ref{tab:pctl}) which means that this altitude is quite typical in the tropics. In the analysis of diurnal cycle and monthly distribution in figure \ref{fig:alt} it can be seen that there is a peak at noon which correspond to maximum solar irradiance. \citet{Williams2006} pointed out that TGFs  produced high in the atmosphere experience less atmospheric attenuation and can be more easily detected at satellite altitudes. The results obtained in this analysis shows clearly that high altitude thunderstorms are quite typical in the tropics where TGFs are more detected. If CTA was critical on TGF production more TGF would be expected due that most of thunderstorms reaches altitudes \( \sim \)16 km. We cannot confirm that altitude is primordial for TGF detection. However it seems an important factor (note that almost all TGFs are detected when CTA over 14 km) but together with high CAPE and lightning activity.

\subsubsection{CTC}

For the CTC it can be seen in table \ref{tab:pctl} that \(75 \% \) of the detected TGFs occur when high cloud tops with temperatures below \(-70^{\circ}\)C cover more than 26000 km\(^2\). These cases represent only \(10 \% \) (table \ref{tab:pctl2}) of the total AGILE passages non detecting TGFs analyzed, so TGFs are detected under not typical conditions. Moreover, these cases occurs more often in the afternoon and during March to May and September to December periods (figure \ref{fig:area}). These distributions are coincident with TGF diurnal cycle and monthly distribution (figures \ref{fig:genhh} and \ref{fig:genmm}). \\

\subsubsection{Summary}

It is shown that \(75 \% \) TGFs detected by AGILE over South America in the period from March 2009 to July 2012 prefer CAPE \(>\) 2588 J/kg, stroke rates \(>\) 80 strokes/minute, number of storms \(>\) 9  and CTC \(>\) 26000 km\(^2\) (table \ref{tab:pctl}). These values are typical of tropical thunderstorms slightly displaced from ITCZ (\citep{Cooray2003},Chapter 1) which is consistent with seasonal migration of AGILE TGFs detected over SA (figures \ref{fig:genhh} and \ref{fig:genmm}) and AGILE pases non detecting TGFs (figures \ref{fig:str} - \ref{fig:area}) as well. These high CAPE values \(>\) 2000J/kg are related to vigorous updrafts which are important for charge separation in mixed phase, suggesting that strong electrification of mixed phase higher up in the cloud (elevated charge mechanism \citep{MacGorman1989}) plays an important role on TGF production. Moreover, more than \(50 \% \) of TGFs were detected under CAPE values above 3400 J/kg (table \ref{tab:pctl}, that are not very typical conditions even in the tropics). The fact that TGFs are detected above large number of strokes and storms may be explained by greater chance that one of the thunderstorms is in a stage of development favourable to produce a TGF. Large flash rates also increases the probability to detect a TGF. Finally, the altitude of thunderstorms seems to be important but not promordial for TGF production because \(50 \% \) of TGFs were detected above CTA \(>\) 16 km (table \ref{tab:pctl}) but it is a quite typical situation in the tropics, comprising \( 60 \% \) of the AGILE passages without detected TGFs analysed.\\

The reason for the small number of TGFs detected outside the tropics would be due to all these conditions that seem to be preferred for TGF production and that are very rare outside tropical regions, together with analysis presented before showing that the same AGILE satellite over Europe would need \(\sim 113\) days of exposure to observe a TGF.

\section{Conclusions}
This paper presents a detailed study of the meteorological conditions conducive to TGF production over South America.
The global analysis of RHESSI and AGILE TGFs and WWLLN strokes in 2009-2012 period has revealed:
\begin{enumerate}
\item A new weak TGF production region over the tropical Pacific.

\item The TGF/lightning ratio for AGILE TGFs and RHESSI TGFs is different than expected compared to \citet{Fuschino2011} for AGILE and \citet{Smith2010} for RHESSI being smaller in America than in Africa and Asia. However these ratios are calculated with a different method and refers mostly to cloud to ground strokes.

\item Geographical occurrrence of TGFs in the three main lightning chimneys follows very well the ITCZ seasonal movement as previously reported \citep{Splitt2010}.

\item TGFs exhibit diurnal cycles in all continental regions higher in afternoon suggesting that solar irradiance and consequently CAPE and convective thunderstorms help to get the right conditions for TGF production.

\end{enumerate}

For the analysis of AGILE passages detecting and non detecting TGFs in the 2009-2012 period over South America we can conclude that:
\begin{enumerate}

\item AGILE TGFs are detected under rare conditions of flash rate, number of storms, CAPE and CTC. These conditions occurs more often in the afternoons and months for which the ITCZ is crossing AGILE coverage region.

\item Higher CAPE values suggests that electrification processes related to vigorous updrafts are important for production of lightning associared to TGFs.

\item High number of thunderstorms below the AGILE satellite increases the probability to have a thunderstorm in the most favorable stage of its life cycle (if there is any) for TGF production. Also, a high number of flashes below the AGILE satellite increases the probability to detect a TGF. 

\item Cloud Top Altitude \citep{Williams2006} is not the main parameter for TGF detection at satellite altitude. However it seems important together with high CAPE and lightning activity.

\item Preferences for TGF production and comparison of South America and Europe annual stroke densities shows that TGF detection in mid-latitudes would be quite difficult with a detector equivalent to MCAL. It would need \(\sim 113\) days of exposure to observe a TGF.
\end{enumerate}

%----------------------------------------------------------------------------------------
%	ACKNOWLEDGEMENTS
%----------------------------------------------------------------------------------------

\section*{Acknowledgments}
This study was supported by research grants from the Spanish Ministry of Economy and Competitiveness (MINECO) AYA2011-29936-C05-04 and ESP2013-48032-C5-3-R. 
We would like to thank E. Williams (Parsons Laboratory, Massachusetts Institute of Technology) for his suggestions and Thunderstorm Effects on Atmosphere and Ionosphere System (TEA-IS) of European Science Foundation (ESF) for funding travel and accommodation for the collaboration between investigation teams.
This study was also supported by the European Research Council under the European Unions Seventh Frame- work Programme (FP7/2007-2013)/ERC grant agreement 320839 and the Research Council of Norway under contracts 208028/F50, 216872/F50, and 223252/F50 (CoE).

%----------------------------------------------------------------------------------------
%	BIBLIOGRAPHY
%----------------------------------------------------------------------------------------

%\bibliographystyle{model2-names}\biboptions{authoryear}
%\bibliographystyle{elsarticle-harv}
%\bibliography{C:/Users/Ferran/Documents/Ferran_Latex/Bib/library2}

%----------------------------------------------------------------------------------------
%	FIGURES
%----------------------------------------------------------------------------------------

%----------------------------------------------------------------------------------------
%	TABLES
%----------------------------------------------------------------------------------------

\clearpage

\begin{table}
\caption{Percentiles 10, 25, 50, 75 and 90 of the five meteorological parameters analyzed of AGILE passages detecting and non detecting TGFs (non detecting TGFs cases in parenthesis)}
%\tablenotemark{a}}
\centering
\begin{tabular}{c | c c c c c}
\cline{2-6}
 &  \multicolumn{5}{ |c| }{\textbf{TGF (Non TGF)}}\\
\hline
\multicolumn{1}{ |c| }{\textbf{Percentile}} & \textbf{\# Strokes}  & \textbf{\# Storms} & \textbf{CAPE (J/kg)} &  \textbf{CTA (km)} & \multicolumn{1}{ c| }{\textbf{CTC (km\(^{2}\))}} \\
\hline
\multicolumn{1}{ |c| } { \textbf{ 10 \% }}   & 12 (2)  & 5(2)  & 0(0) & 14(13)& \multicolumn{1}{ c| }{21304(232)}  \\ 
\multicolumn{1}{ |c| }{ \textbf{ 25 \% }}  & 27 (5)  & 9(3)  & 2588(1606) & 15(13)& \multicolumn{1}{ c| }{26746(639)}  \\
\multicolumn{1}{ |c| }{ \textbf{ 50 \% }}   & 59 (13)  & 14(5)  & 3444(2371) & 16(16) & \multicolumn{1}{ c| }{46774(2749)}  \\
\multicolumn{1}{ |c| }{ \textbf{ 75 \% }}   & 205 (34)  & 20(8)  & 4008(3197) & 17(17) & \multicolumn{1}{ c| }{57422(10667)}  \\
\multicolumn{1}{ |c| }{ \textbf{ 90 \% }}   & 277(70)  & 26(13)  & 4651(4064) & 17(17) &  \multicolumn{1}{ c| }{91972(27654)}  \\
\hline
\end{tabular}
%\tablenotetext{a}{Footnote text here.}
\label{tab:pctl}
\end{table}

\begin{table}
\caption{Percentiles of the five meteorological parameters analyzed of the AGILE passages non detecting TGFs corresponding to the percentiles values 10, 25, 50, 75 and 90 of AGILE passages detecting TGFs}
%\tablenotemark{a}}
\centering
\begin{tabular}{l | c c c c c}
\cline{2-6}
 &  \multicolumn{5}{ |c| }{\textbf{Non TGF Percentiles}}\\
\hline
 \multicolumn{1}{ |c| }{\textbf{TGF Percentile}} & \textbf{\# Strokes}  & \textbf{\# Storms} & \textbf{CAPE} & \textbf{CTA} & \multicolumn{1}{ c| }{\textbf{CTC}} \\
\hline
\multicolumn{1}{ |c| }{ \textbf{ 10 \% }}  & \( 49 \% \)  & \( 46 \% \) & \( 0 \% \) & \( 31 \% \) & \multicolumn{1}{ c| }{\( 87 \% \)}  \\ 
\multicolumn{1}{ |c| }{ \textbf{ 25 \% }}  & \( 69 \% \)  & \( 76 \% \) & \( 57 \% \) & \( 38 \% \) & \multicolumn{1}{ c| }{\( 90 \% \)}  \\
\multicolumn{1}{ |c| }{ \textbf{ 50 \% }}  & \( 87 \% \)  & \( 91 \% \) & \( 81 \% \) & \( 40 \% \) & \multicolumn{1}{ c| }{\( 97 \% \)}  \\
\multicolumn{1}{ |c| }{ \textbf{ 75 \% }} & \( 99 \% \)  & \( 98 \% \) & \( 89 \% \) & \( 64 \% \) &\multicolumn{1}{ c| }{ \( 97 \% \)} \\
\multicolumn{1}{ |c| }{ \textbf{ 90 \% }}  & \( 99 \% \)  & \( 99 \% \) & \( 95 \% \) & \( 64 \% \) & \multicolumn{1}{ c| }{\( 99 \% \)}  \\
\hline
\end{tabular}
%\tablenotetext{a}{Footnote text here.}
\label{tab:pctl2}
\end{table}

\end{document}